\documentclass[times, twoside, watermark]{Arxiv-Spectroscopy-ASM}
\usepackage{blindtext}

\leadauthor{John M. Bass} 

\begin{document}

\title{An Angular Spectrum Approach to Inverse Synthesis for the
Characterization of \\Optical and Geometrical Properties of Semiconductor Thin Films}
\shorttitle{}

\author[1,*]{John M. Bass}
\author[2]{Manuel Ballester}
\author[3]{Susana M. Fernández}
\author[2,4]{\\Aggelos K. Katsaggelos}
\author[5]{Emilio Márquez}
\author[1,2,4]{Florian Willomitzer}

\affil[1]{Wyant College of Optical Sciences, University of Arizona, Tucson, AZ 85721, United States}
\affil[2]{Department of Computer Science, Northwestern University, Evanston, IL 60208, United States}
\affil[3]{Department of Energy, CIEMAT, Avenida Complutense 40, 28040 Madrid, Spain}
\affil[4]{Department of Electrical and Computer Engineering, Northwestern University, Evanston, IL 60208, United States}
\affil[5]{Department of Condensed-Matter Physics, University of Cadiz, 1510 Puerto Real, Spain \vspace{1em}}
\affil[*]{Correspondence: jmbass\at arizona.edu, fwillomitzer\at arizona.edu}

\maketitle

\begin{abstract}
To design semiconductor-based optical devices, the optical properties of the used semiconductor materials must be precisely measured over a large band. Transmission spectroscopy stands out as an inexpensive and widely available method for this measurement but requires model assumptions and reconstruction algorithms to convert the measured transmittance spectra into optical properties of the thin films. Amongst the different reconstruction techniques, inverse synthesis methods generally provide high precision but rely on rigid analytical models of a thin film system. In this paper, we demonstrate a novel flexible inverse synthesis method that uses angular spectrum wave propagation and does not rely on rigid model assumptions. Amongst other evaluated parameters, our algorithm is capable of evaluating the geometrical properties of thin film surfaces, which reduces the variance caused by inverse synthesis optimization routines and significantly improves measurement precision. The proposed method could potentially allow for the characterization of ``uncommon'' thin film samples that do not fit the current model assumptions, as well as the characterization of samples with higher complexity, e.g., multi-layer systems.
\end {abstract}

\section{Introduction}
Semiconductor-based optical devices, such as light sensing devices or solar panels, must be designed and manufactured with precise knowledge of the semiconductor optical properties in order to attain maximum performance \cite{Chayma_2020, Rodriguez-Tapiador_2023, lopezenhanced, marquez2023complex}. Of particular importance are two optical constants: The refractive index \textit{n}, and the extinction coefficient $\kappa$. These constants form the complex refractive index $\Tilde{n} = n + \iu{\kappa}$, which defines how light propagates through and gets absorbed by a medium. Precise knowledge of the complex refractive index is thus important for semiconductor research and design - not at least because the complex refractive index can also be expressed as a multivariate function of parameters essential in diverse industrial applications \cite{petrik2012optical}. This includes the description of electronic transitions, which are instrumental in determining, e.g., the optical bandgap energy $E_\text{g}$, a critical factor for the functionality of optical sensors \cite{islam2020band, panda2012preparation, nesheva2019changes}. Furthermore, it facilitates the assessment of molecular structure disorder through the Urbach energy $E_\text{u}$, which is particularly valuable for detecting defects in crystalline materials \cite{marquez2023optical, kumar2014evolution} or gauging the level of disorder in amorphous materials \cite{ballester2022energy, amato2020observation}, all key indicators of the quality of sensor materials. 

\begin{figure*}[ht!]
\centering\includegraphics[width=0.8\linewidth]{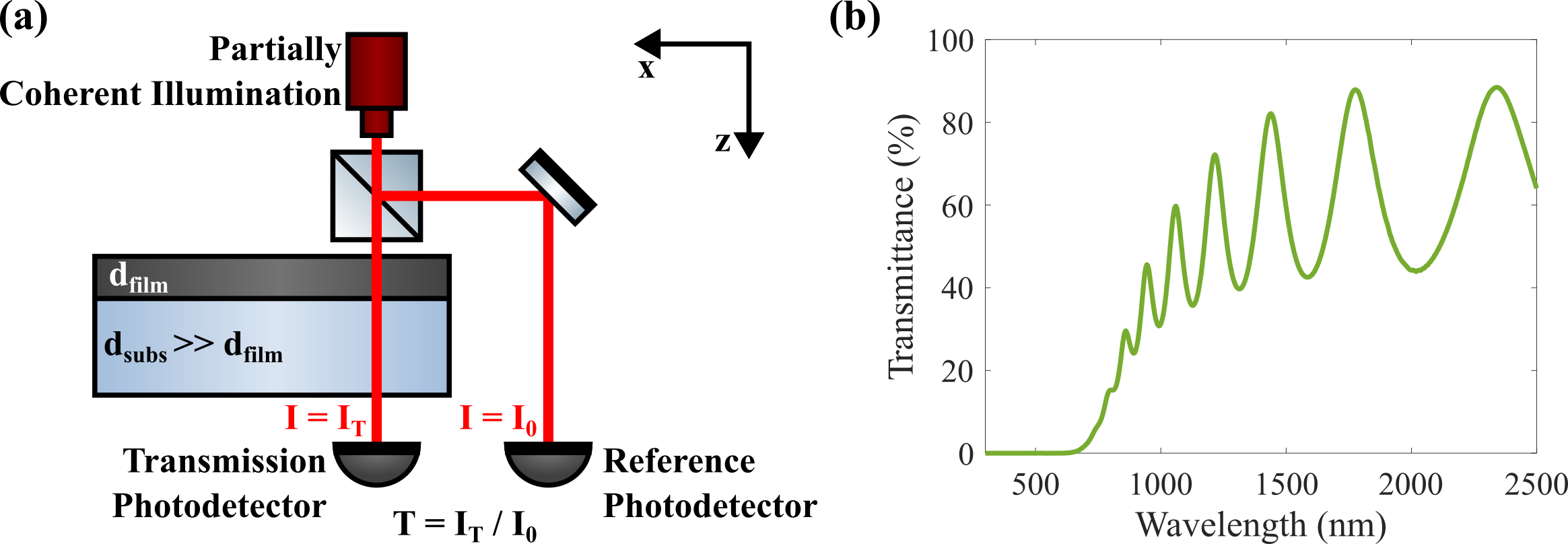}
\caption{(a) Standard double-beam spectrophotometer schematic: A partially coherent illumination beam is split, with one path incident upon a thin film sample that has been deposited on a glass substrate. Both beam paths are then recorded by photodetectors, and the transmittance of the thin film-substrate system is computed from the measurements.  (b) Example of a measured thin-film transmittance spectrum, taken from the sample that was previously characterized and discussed in \cite{Ballester_2022_Application}. The oscillations in this spectrum are caused by interferences between reflected beams within the thin film, similar to what happens in a Fabry-Perot interferometer~\cite{Kreis_2022_Foundations}. The transmittance spectrum can be converted to optical property spectra using a variety of computational algorithms.}
\label{fig:SpectrophotometerSchematic}
\end{figure*} 

While refractive index and extinction coefficient are often called ``optical constants'' in literature (including this paper), we emphasize that they are wavelength-dependent, meaning that it becomes necessary to measure these properties over a broad range of frequencies.
For a typical measurement, a thin film sample of the semiconductor material is prepared and deposited on a glass substrate. From there, the optical properties of the thin film can be found using a variety of methods. A frequently used method is ellipsometry \cite{Thompkins_2005}, which characterizes a material by the change in polarization of reflected light. Ellipsometry is a well-developed method that has been widely explored in the literature \cite{fujiwara2007spectroscopic, kalas2017ellipsometric, romanenko2020membrane, lohner2020determination, marquez2023mid}. However, devices that can perform ellipsometric measurements are high-cost, and their availability is often limited due to the complexity of the method. In contrast, transmission spectroscopy \cite{Heavens_1991} stands out as a high-quality alternative that characterizes a material by analyzing the transmittance spectrum of the thin film. Since this procedure does not rely on polarization measurements and operates at a single normal-incidence angle, it bears a lower system complexity and cost than ellipsometry. Transmission spectrophotometers are already widely available \cite{Bosch_1998} and are specifically appropriate for low-to-medium absorption environments where most of the incidence light will pass through the thin-film sample \cite{van2005determination, saleh2017evaluation, minkov2020perfecting}. Under these conditions, transmission-based measurements incorporate information about the full volume of a semiconductor and are less affected by surface conditions of the films (e.g., surface roughness) compared to reflection-based measurements like ellipsometry.

A diagram of a typical double-beam transmission spectrophotometer is shown in Fig. \ref{fig:SpectrophotometerSchematic}(a). During the measurement, a partially coherent beam illuminates a thin film sample under test. A common spectral linewidth for the respective illumination system is 2 nm, though modern spectrophotometers can have linewidths ranging between 0.1 nm to 10 nm  \cite{Hedderich_2023, PerkinElmer_2016}. The thin film typically has a thickness in the micrometer range, while the substrate has a typical thickness on the order of a few millimeters. The illumination sweeps through a broad band of wavelengths in the UV-VIS-NIR spectral range, typically ranging between 300 nm and 2500 nm \cite{Hedderich_2023, PerkinElmer_2016}, while a detector measures the transmitted power for each wavelength. During the illumination sweep, a reference detector simultaneously measures the beam power to record any intensity fluctuations that may have occurred in the illumination. After a second reference measurement of the glass substrate alone,  the transmittance spectrum of the thin film system is obtained. An example of a typical spectrum is shown in Fig \ref{fig:SpectrophotometerSchematic}(b). Eventually, the optical constants can be calculated from the spectrum via different algorithmic procedures.

 A frequently used approach is the Swanepoel method \cite{Manifacier_1976, Swanepoel_1983, Swanepoel_1984, jin2017improvement, marquez2019influence}, which calculates the spectra for \textit{n} and \textit{k} from the envelope of the transmittance spectrum. While this method is fast and works with little computational effort \cite{tejada2019determination}, it fits \textit{n} and $\kappa$ only from a set of specific wavelengths that correspond to the tangential points where the spectrum intersects with its envelope. However, due to the complexity of evaluating tangential points and envelope  \cite{Ballester_2022_Comparison}, and the used approximation that \(\kappa^2 \ll n^2\), the Swanepoel method typically delivers lower accuracy results compared to inverse synthesis methods.

The basic idea of inverse synthesis methods \cite{Dobrowolski_1983, Poelman_2003, ventura2005optimization} is to apply algorithmic optimization routines on analytical models of thin film systems to ``reverse-engineer'' the optical properties of a thin film. Many techniques rely on transfer matrix approaches, which offer accurate simulation results and fast computation times \cite{Knittl_1976, yeh2006optical, prentice2000coherent}. However, the employed analytical models typically have rigid assumptions about the sample, which limits their applicability only to conventional and ``well-prepared'' samples. For instance, while some models correct for a tilted top surface of the thin film layer~\cite{Ruiz_2020}, accounting for higher-order shape variations of the top surface is often overlooked in most state-of-the-art models. In addition, the inclusion of potential beam tilt into these models is rare, and sample inhomogeneity (e.g., refractive index variations), as well as the spectral or spatial distributions of incident beams, are typically not considered. For all these cases, existing models often produce poor fits, leading to incorrect characterizations.

In this paper, we introduce a novel inverse synthesis method that does not rely on strong model assumptions and hence can account for a large variety of potential error sources. We demonstrate fit qualities on par with current state-of-the-art methods while simultaneously maintaining great flexibility and accounting for various conditions that could potentially lead to improved characterization of semiconductor devices in the case of unusually prepared or exotic samples.

\section{Methods}
As discussed, the goal of this paper is to improve the versatility of thin film transmission spectroscopy systems by a new inverse synthesis method that can provide precise characterization of thin film semiconductor samples, even in the presence of potential material imperfections created during the sample preparation process. Our novel method is divided in two ``building blocks''. The first block is a novel simulation engine based on the split-step angular spectrum method. The simulator takes an arbitrary given system geometry and configuration and an arbitrary set of sample parameters within reasonable boundaries to accurately simulate light propagation through the respective media and to model the respective transmittance spectra.
The simulation engine is embedded into the second building block: a global optimizer. The optimizer takes a measured thin film spectrum and attempts to find the most likely sample properties that would reproduce the measurement in our simulator via a genetic algorithm \cite{Alhijawi_2024_Genetic}. 
The so-obtained optical properties accurately characterize the optical system. In the following subsections, we will describe the design of the simulator and optimizer in detail.

\subsection{Simulation of light propagation through a thin film system}
\begin{figure}[b!]
\centering\includegraphics[width=\linewidth]{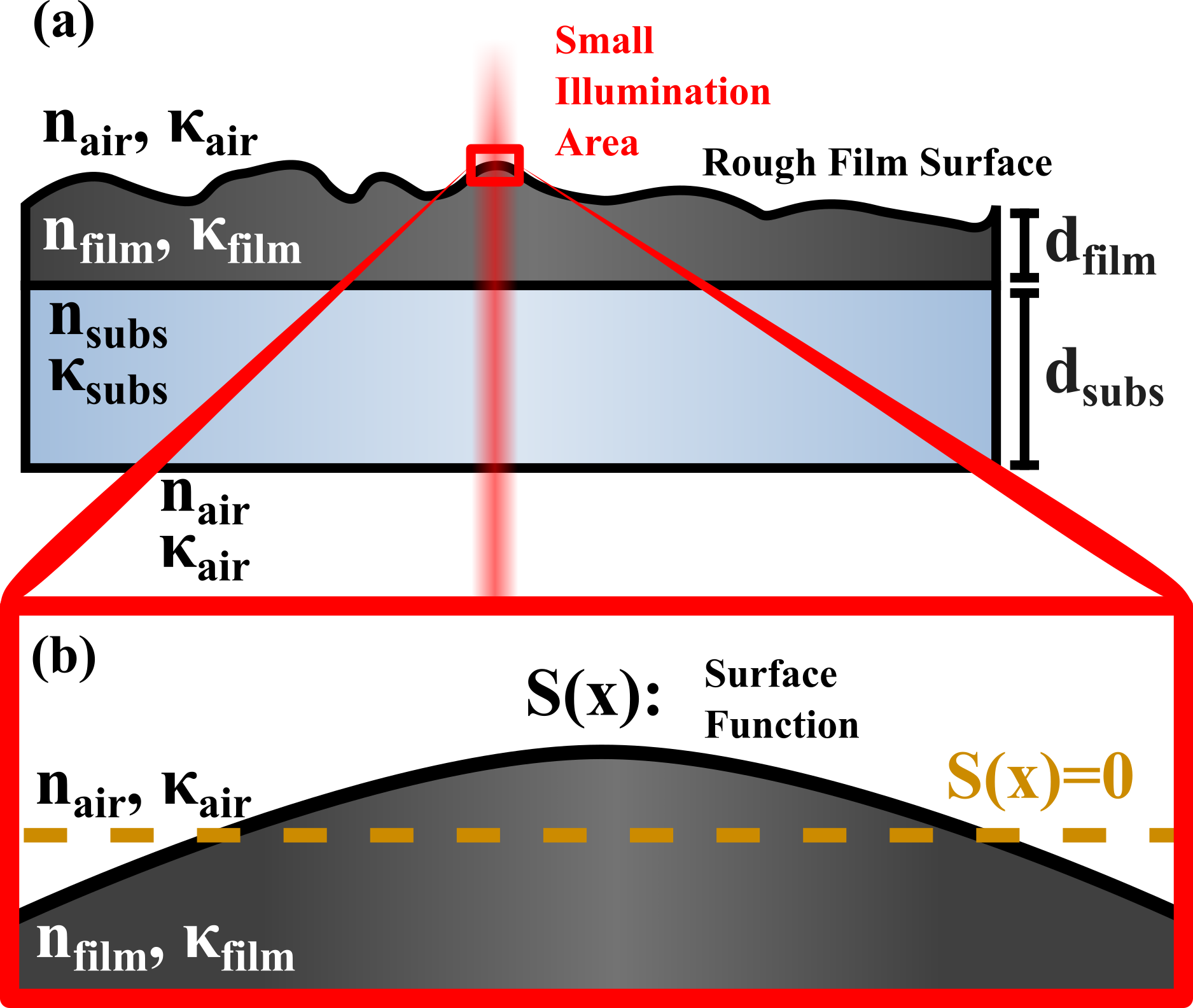}
\caption{(a) Representation of a thin film system in our angular spectrum propagation simulator. An illumination propagates through a small amount of air, before landing upon a thin film semiconductor. Then, after propagating through the semiconductor, it travels through a glass substrate before entering open air again. Although any medium is capable of having a variable surface curvature on either side,  we only vary the top film surface in this paper to optimize runtime. (b) Within the small illumination area, the thin film surface is  modeled as a smooth surface with variations defined by a zero-mean function S(x). }
\label{fig:SurfaceFunctionDiagram}
\end{figure}

To accurately simulate a thin film system, we used a variation of the Angular Spectrum Method \cite{Goodman_1969}, which we modified to allow for non-planar interfaces \cite{Bass_2024_Angular, Bass_2024_Increasing}. We represent a thin film system as a layer of four volumes: an air volume, followed by the thin film sample on a thick glass substrate, followed by another air volume (see Fig. \ref{fig:SurfaceFunctionDiagram}(a)). The top surface of the thin film is parameterized to allow for tilt and curvature, while every other surface in the system is, without limiting generality, assumed to be perpendicular to the optical axis of the detector. 

The angular spectrum  method is a Fourier optics approach to simulate how light propagates and diffracts through an environment. The method is precise, as it is an equivalent solution to the Rayleigh-Sommerfeld integral equations, and strictly satisfies the Helmholtz equation \cite{Sherman_1967_Application}. We briefly introduce the method in the following and direct the reader to \cite{Goodman_1969} for a more elaborate description. 

If light is propagating in the +z direction (see Fig. \ref{fig:SpectrophotometerSchematic}(a)), a 1D cross section of the propagating field at \(z=z_{0}\) can be represented as \(E(x,z_{0})\). The Fourier transform of this cross-section, \(F_{E}(f_{x},z_{0})\), can be interpreted as a series of amplitudes of tilted plane waves, which interfere to produce \(E(x,z_{0})\). To propagate the field forward, each component plane wave is propagated to a new plane, \(z=z_{0} + \Delta z\), by applying the phase \( \Delta \phi = k_{z}\Delta{z}\) to each component.  Then, the resulting superposition of plane waves produces the field at the new plane, \(E(x,z_{0} + \Delta{}z)\). The longitudinal wavenumber \(k_{z}\)  can be computed from the transverse wavenumber \(k_{x}\), which can be defined in terms of the spatial frequency of a component plane wave.
\begin{equation}\label{eq:TransverseWavenumberDefinition}
k_{z} = \sqrt{k^{2} - k_{x}^{2}} = \frac{2\pi{}}{\lambda{}}\sqrt{n^{2} - \lambda{}^{2}f_{x}^{2}}
\end{equation}
Using this definition, a free space transfer function, which propagates an electric field a distance \(\Delta{}z\), can be defined 
\begin{equation}\label{eq:InhomogenousElectricFieldKernel}
P_{AS}(f_{x},\Delta{}z) = \exp{\left(\iu{}k\Delta{}z\sqrt{n^{2}-\lambda{}^{2}f_{x}^{2}} \right)}~~,
\end{equation}
and applied to \(F_{E}(f_{x},z_{0})\) for propagation:

\begin{equation}\label{eq:KernelMultiplication}
F_{E}(f_{x},z_0+\Delta{}z) = F_{E}(f_{x},z_0)P_{AS}(f_{x},\Delta{}z)
\end{equation}

\begin{figure}[b!]
\centering\includegraphics[width=0.7\linewidth]{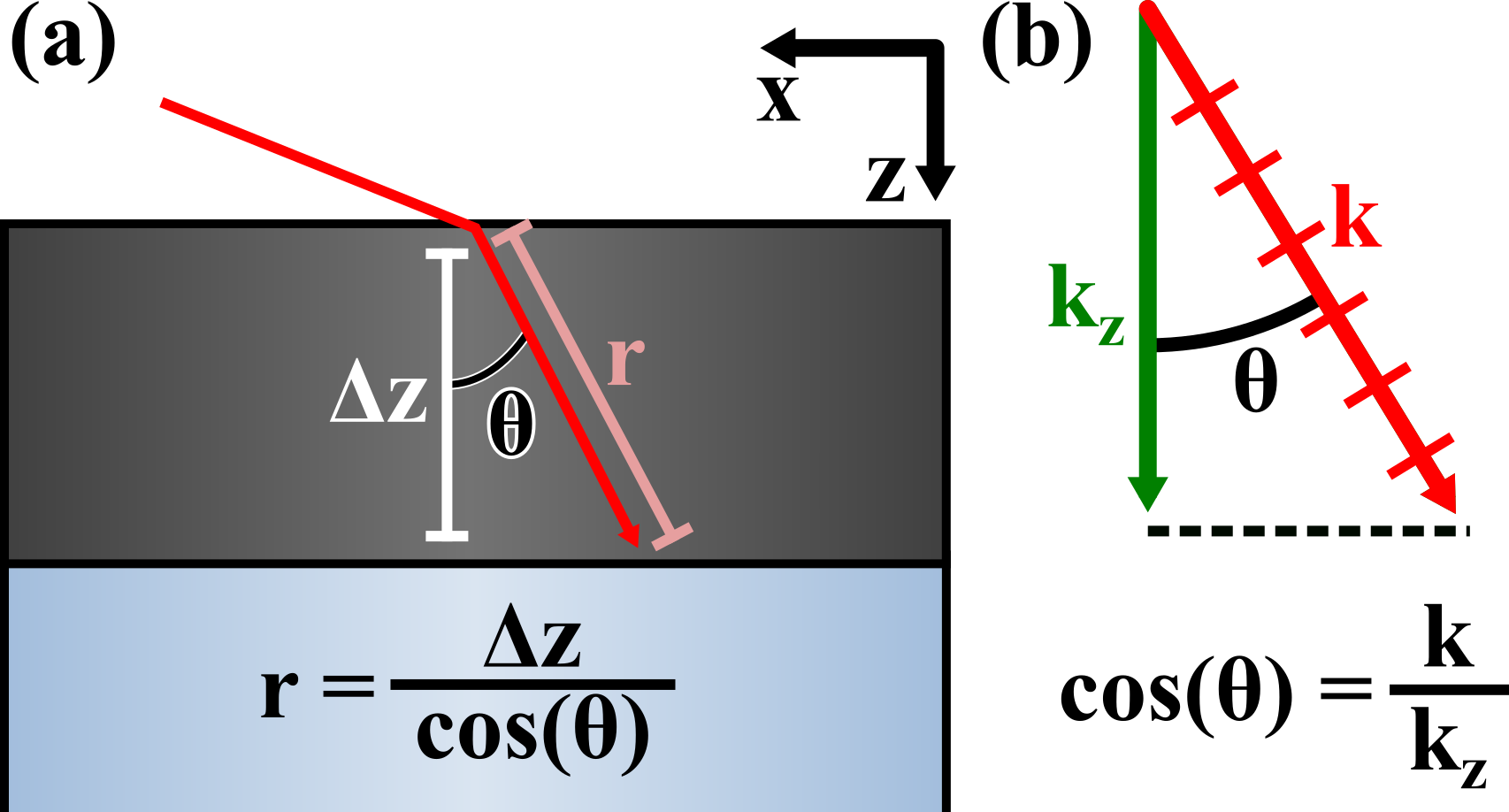}
\caption{(a) Propagation distance of an angled plane wave component in a homogenous medium. (b) Calculation of plane wave angle in angular spectrum domain.}
\label{fig:AngledAbsorptionDiagram}
\end{figure}

The ``standard'' angular spectrum method described above can be used to propagate between homogeneous, non-absorptive planar media with any refractive index, without approximations beyond the Helmholtz equation. Since the thin film absorbs light, however, the absorption must also be accounted for. One method of simulating absorption with an angular spectrum method was demonstrated for ultrasonic waves in \cite{vyas_2012}, in which the authors calculated an amount of absorption based on the average ray angle of the propagated wavefront. This approach, however, does not account for the fact that angular spectrum plane wave components with high spatial frequencies have a longer propagation distance inside the thin film than a normal-incidence plane wave, due to their angle of travel. This causes high-spatial-frequency components to be absorbed slightly more than low-spatial-frequency components, creating a slight beam-widening effect. To consider this effect, we generalized the standard angular spectrum kernel to account for the unique absorption of every component plane wave. According to Beer's law and the geometry \cite{vyas_2012} shown in Fig. \ref{fig:AngledAbsorptionDiagram}, a plane wave that propagates through a medium along $r= \Delta z / \cos{\theta}$ would be attenuated by 

\begin{equation}\label{eq:ComponentPlaneWaveAbsorption}
E_{out} = E_{in} \exp{\left(-\frac{2\pi{}}{\lambda{}} \kappa r\right)} = E_{in} \exp{\left(-\frac{2\pi{}}{\lambda{}}\frac{ \kappa \Delta z}{\cos{\theta}}\right)}~.
\end{equation}

\begin{figure*}[t!]
\centering\includegraphics[width=15cm]{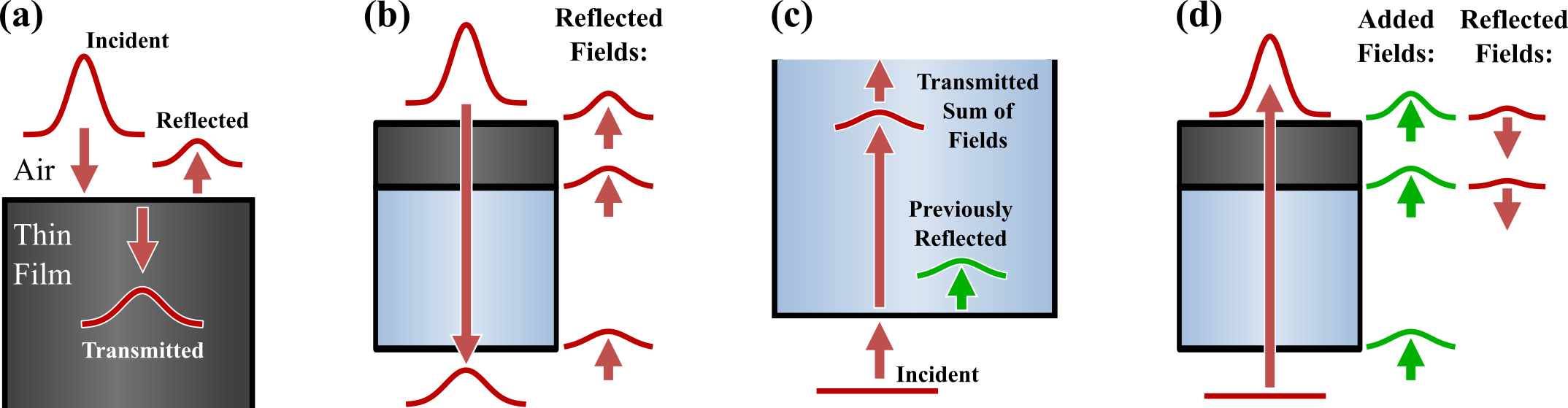}
\caption{Split-Step Angular Spectrum method for multi-layered systems: (a) Our model starts by propagating to the (potentially curved) thin film surface, and computes the transmitted and reflected fields. (b) It propagates the transmitted field through every interface in the system, saving reflected fields. (c) To start a reverse propagation step, the model propagates an empty field (zero amplitude), then adds a previously saved reflected field at the first interface. (d) The model finishes a reverse propagation step by repeating this process through every interface in the system, saving reflected fields for the next forward propagation step.}
\label{fig:ForwardBackwardPropagation}
\end{figure*} 

\noindent By replacing the angular component of this absorption formula with the ratio of wavenumbers $\cos{\theta} = k / k_{z} = \frac{1}{n} \sqrt{n^{2}-\lambda{}^{2}f_{x}^{2}}$, we can now define  a transfer function which describes the absorption of component plane waves in the angular spectrum

\begin{equation}\label{eq:AbsorptionTransferFunction}
P_{abs}(f_{x},\Delta{}z) = \exp{\left(-\frac{2\pi{}}{\lambda{}}\frac{ \kappa n \Delta z}{\sqrt{n^{2}-\lambda{}^{2}f_{x}^{2}}}\right)} ~~, 
\end{equation}  
which  is then combined with the standard angular spectrum kernel to produce a complex kernel that models both propagation and absorption:
\begin{equation}\label{eq:CompleteTransferFunction}
P_{\text{total}}(f_{x},\Delta{}z) = P_{AS}(f_{x},\Delta{}z) P_{abs}(f_{x},\Delta{}z)
\end{equation}

This adjusted transfer function allows a complex field to be propagated through a homogenous material with minimal approximations. To propagate through a multi-layered medium such as our thin film system, however, additional modifications must be made to the simulator. For this, we use  a split-step approach that is inspired by the Hybrid Angular Spectrum Method \cite{vyas_2012}, and other angular spectrum approaches such as~\cite{Xu_2022, Mellin_2001}:  We assume that each successive medium is homogeneous within its volume, but one or more boundaries might not be perfectly planar, as shown in Fig. \ref{fig:SurfaceFunctionDiagram}(a).

To propagate through such a system, we start at the first medium - a small volume of air. After propagating through the medium with the angular spectrum method, we arrive at the (potentially curved) boundary of the thin film where the reflected and transmitted fields are calculated. This is done by multiplying the incident field with a modified reflection/transmission coefficient which consists of the standard Fresnel reflection/transmission coefficients at normal incidence \(r_F\) and \(t_F\), and additional ``shape-based'' coefficients \(r_{\Delta{}}\) and \(t_{\Delta{}}\) which account for the varying surface shape described by  $S(x)$.

\begin{equation}\label{eq:ReflectedElectricField}
E_\text{reflected}(x,z) = r_{F}r_{\Delta{}}E(x,z)
\end{equation}
\begin{equation}\label{eq:TransmittedElectricField}
E_\text{transmitted}(x,z) = t_{F}t_{\Delta{}}E(x,z)
\end{equation}

\noindent
Here, the Fresnel coefficients are given by 

\begin{equation}\label{eq:FresnelReflectionCoefficient}
r_{F} = \frac{\Tilde{n}_2 - \Tilde{n}_1}{\Tilde{n}_2 + \Tilde{n}_1}~~, ~~ t_{F} = \frac{2\Tilde{n}_1}{\Tilde{n}_2 + \Tilde{n}_1}~~ ,
\end{equation}
with ``medium 1'' being air and ``medium 2'' being the semiconductor material in this case. As shown and explained in Appendix \ref{appendix:coefficients}, the shape-based coefficients have been derived  to

\begin{subequations}\label{eq:ShapeBased}
\begin{align}
r_{\Delta{}}(x) &= \exp{\left(-a_{\Delta{}r}(x) + \iu{}\phi{}_{\Delta{}r}(x)\right)} \tag{\ref*{eq:ShapeBased}.a} \\
t_{\Delta{}}(x) &= \exp{\left(-a_{\Delta{}t}(x) + \iu{}\phi{}_{\Delta{}t}(x)\right)} ~~.\tag{\ref*{eq:ShapeBased}.b}
\end{align}
\end{subequations}

Applying both the Fresnel and ``shape-based'' reflection and transmission coefficients allows accurate simulation of the behavior of electric fields at a deformed boundary, and applying the angular spectrum method allows for fields to be propagated between boundaries. Thus, by combining these two principles, we can now simulate the forward propagation of light through a layered thin film system, as shown in Fig. \ref{fig:ForwardBackwardPropagation}(a) and (b). However, with a single forward propagation, the results will not be entirely accurate, as reflected beams will back-reflect and add to the transmitted field. To account for this effect, we apply additional modifications to our simulator: We save the reflected fields at every interface, then perform a reverse propagation step, in which we start with a zero-intensity field, but add reflected fields at every interface before continuing propagation. This reverse propagation step is depicted in Fig. \ref{fig:ForwardBackwardPropagation}(c) and (d). During the reverse propagation step, we save the newly reflected fields so that they can be propagated in the next forward propagation step (Fig. \ref{fig:ForwardBackwardPropagation}(d)). We repeat the process of simulating forward and reverse propagations until the net field being propagated has attenuated to a total field strength of less than 30dB of the original value. Finally, we sum the net transmitted and reflected fields from every propagation to obtain physically accurate electric field distributions that would be produced by such a system.

To accurately approximate the partial coherence properties of the spectrophotometer illumination, we simulated the transmittance of many wavelengths over the whole spectrum, then selectively scaled and summed the resultant electric fields to obtain the interference pattern that would be obtained by a partially coherent beam with a uniform spectral distribution at a desired center wavelength. This simulation was repeated for many center wavelengths, to obtain many transmitted electric fields which densely sample our generated transmittance spectrum. For all our simulations, we assumed a linewidth of $4 nm$. To maximize computation speed, we stop each simulation when the field at each center wavelength has attenuated beyond the 30dB cutoff. Each transmitted electric field distribution is then used to calculate the total transmitted power and the net transmittance of the thin film system for the simulated wavelength. For larger center wavelengths, our simulator produces additional etalon effects caused by inter-reflections in the substrate. These effects show up as high-frequency variations in our simulated transmittance spectrum. While physically sound, these variations rarely appear in measured spectra, possibly because of device-internal post-processing. As commonly done in the literature \cite{king1999analysis, zimmermann2013optimizing, zhao2014parameters}, we apply a Savitzky-Golay filter of order 3 and frame length 11 to smooth the spectrum.
An example transmittance spectrum simulated with our full pipeline is shown in Fig. \ref{fig:BestFitPlots}(b). In the next step, the simulated transmittance spectrum is compared to a measured spectrum to generate a figure of merit for the optimization of material and surface parameters. \\

\subsection{Evaluation of  optical properties with a global  optimizer}

\begin{table}[b!]
    \centering
    \begin{tabular}{L{0.21\linewidth} L{0.50\linewidth} L{0.13\linewidth}}
         \hline
         System \newline Parameter & Description & Unit\\
         \hline
         $A$ & Lorentz Oscillation Amplitude & eV \\
         $E_0$ & Peak Transition Energy & eV\\
         $E_\text{G}$ & Bandgap Energy & eV\\
         $C$ & Lorentz Oscillation \newline Broadening Term & eV\\
         $E_\text{C}$ & Urbach Threshold Energy & eV\\
         $\epsilon_{1, \infty} = 1$ & Infinite Frequency \newline Dielectric Function & unitless\\
         $d_\text{film}$ & Film Thickness & $\mu m$\\
         $\theta_\text{in}$ & Beam Incidence Angle & deg \\
         $c_1$ & First-Order Surface Shape \newline Coefficient & unitless\\
         $c_2$ & Second-Order Surface Shape \newline Coefficient & $m^{-1}$ \\
         \hline
    \end{tabular}
    \caption{Parameters used in our method to fully describe the optical and geometrical properties of a measured thin film sample.}
    \label{tab:OptimizationParameters}
\end{table}
To evaluate the optical properties of a given thin film sample with a measured transmittance spectrum,  we incorporate our simulator into a forward model for a global optimizer. As discussed, our simulator takes specific properties of the sample and substrate (like complex refractive index, layer thickness, and surface shape) to generate a full transmittance spectrum. A naive way to evaluate $n$ and $\kappa$ would be to optimize directly for these quantities, i.e., iteratively changing $n$ and $\kappa$ along shape and thickness in the simulator until the simulated spectrum closely resembles the measured spectrum. However, this would mean that $n$ and $\kappa$  would need to be determined for every single wavelength in the spectrum which leads to a very large optimization space and potential overfitting. To circumvent this problem, we utilize an additional material model that is able to generate values for $n$ and $\kappa$ for all wavelengths from only a few parameters.
For the results introduced in this paper, we use the Single-Oscillator Tauch-Lorentz-Urbach (TLU) model developed by Foldyna et al. \cite{Foldyna_2004}, which generates spectra for $n$ and $\kappa$ from only six parameters: The Lorentz Oscillation Amplitude $A$, Peak Transition Energy $E_0$, Bandgap Energy $E_\text{G}$, Lorentz Oscillation Broadening Term $C$, Urbach Threshold Energy $E_\text{C}$, and the infinite-frequency dielectric function $\epsilon_{1, \infty}$. Appendix \ref{appendix:tluModel} further describes the different parameters and their connection to $n$ and $\kappa$. For additional details, we refer to \cite{Foldyna_2004}. Implementing this model reduces the total optimization space to only the 9 parameters shown in Tab. \ref{tab:OptimizationParameters} (note that the coefficient $\epsilon_{1, \infty}$ is fixed to 1, as commonly done in the literature \cite{Foldyna_2004}).

\begin{figure}[t!]
\centering\includegraphics[width=\linewidth]{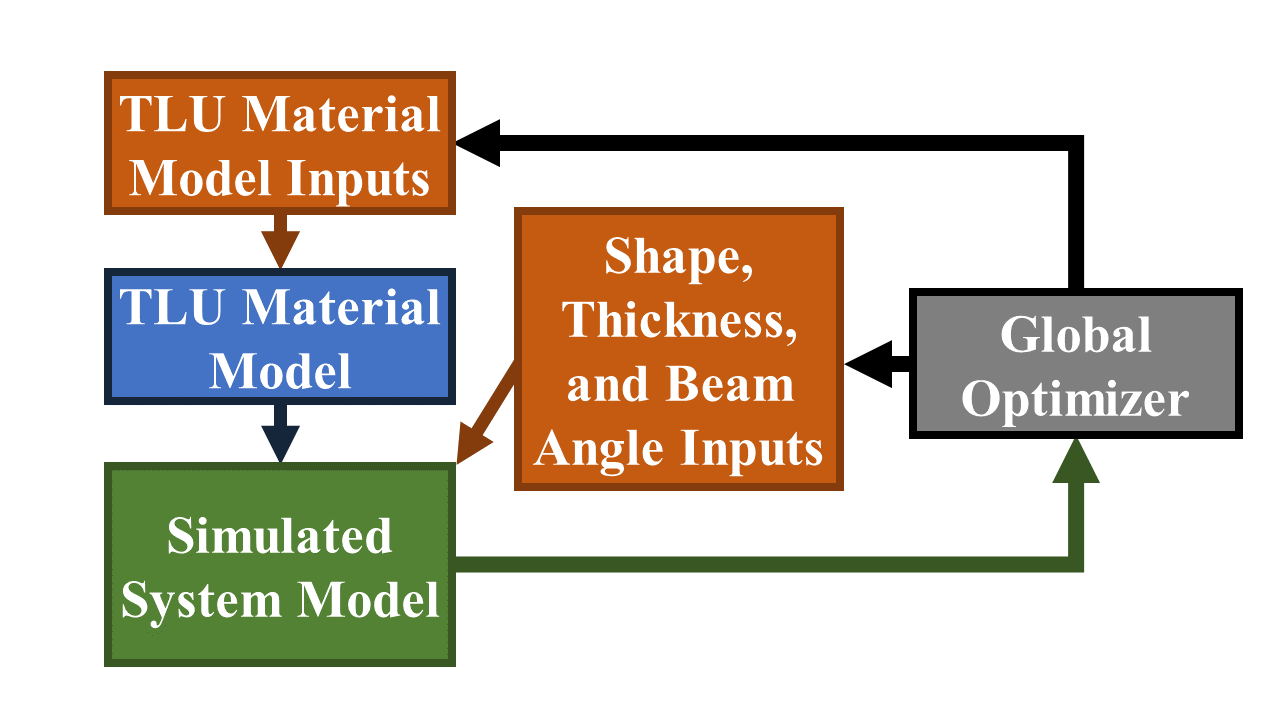}
\caption{Block Diagram of our optimization routine. A global optimizer computes the RMSE between the transmittance spectra of a simulated and measured system, and uses this to compute optimal inputs for a TLU material model, as well as optimal geometrical parameters (film thickness, beam angle, and film surface shape) to fit the measured spectrum and hence the ground truth of the thin film sample as close as possible.}
\label{fig:OptimizationRoutineBlockDiagram}
\end{figure}  

After implementing the material model in our process, our final optimization pipeline is as follows (also see diagram in Fig. \ref{fig:OptimizationRoutineBlockDiagram}):
We use a starting set of ``guessed'' material input parameters to generate the respective values for $n$ and $\kappa$ with the TLU material model \cite{Foldyna_2004}. Eventually, our simulator takes these $n$ and $\kappa$ values together with initial guesses for film thickness $d_\text{film}$, beam incidence angle $\theta_\text{in}$ and surface shape coefficients $c_1$, $c_2$ (see below), to simulate an initial transmittance spectrum. This simulated spectrum is compared with the real measured spectrum by calculating the Root Mean Squared Error (RMSE) between both spectra. The RMSE is then used as a figure of merit by a global optimizer which generates new inputs for the TLU model and simulator via a genetic algorithm~\cite{Alhijawi_2024_Genetic}. The process is repeated until either 200 optimization iterations have passed, or until the best result has remained the same for 30 iterations. After completing a genetic algorithm fit, the best result is fine-tuned using an interior point local optimizer \cite{Anders_2002_Interior, Gondzio_2012_Interior} that uses the same boundaries and inputs as the genetic algorithm.

The work presented in this paper optimizes the surface shape up to the 2nd order (tilt and parabolic curvature), meaning that the shape term is represented by $S(x) = c_1x + c_2x^2$. More complex surface shapes are possible by adding higher order coefficients $c_3, c_4, ... $
to the optimization space, however, at the cost of increased calculation time.

\section{Results}

We validated our method by characterizing transmittance data from an amorphous silicon sample that was prepared using the radio frequency magnetron sputtering technique discussed in~\cite{Fernandez_2021_Sputtered}. This specific sample was previously characterized in \cite{Ballester_2022_Application} with a transfer matrix-based inverse synthesis method \cite{yeh2006optical, prentice2000coherent} using the TLU model from \cite{Foldyna_2004} in combination with
additional surface and thickness measurements from an atomic force microscope. The results of this previous characterization are given in column 2 of Tab. \ref{tab:OptimizationResults}.

To evaluate the sample with our approach, we make additional modifications to speed up the process: During our global optimization, we assume a set of ``rough'' simulator settings to evaluate our parameters. After the optimization is complete, we simulate the transmittance spectrum again with ``finer'' simulator settings to estimate the RMSE between the simulated and measured spectra as accurately as possible.
 Both sets of settings and assumptions are shown in Tab. \ref{tab:OptimizationInputs}. The evaluation results obtained from our method are shown in Tabs. \ref{tab:OptimizationResults} and \ref{tab:OptimizationStatistics} and are explained in the following: To ensure that our method found solutions close to the global minimum and to evaluate the statistical properties of our proposed technique, we performed not only one, but multiple optimization evaluations of the given sample transmittance spectrum. For this we used the University of Arizona's Puma supercomputer, utilizing 10 nodes with 94 CPUs each, and optimizing for 10 hours per evaluation cycle. In total, we ran two 10-hour evaluation cycles. One, where geometry optimization of the thin film surface shape and incidence beam angle was enabled, and one ``comparison cycle'' where the thin film surface was assumed to be flat, and the beam was at normal incidence ($\theta_\text{in} = 0, ~ c_1 = 0,~c_2 = 0$). Within the 10 given hours of each cycle, we obtained 26 fits with enabled geometry optimization, and 30 fits for the cycle without geometry optimization. The optimization parameter bounds for each case are shown in Tab.~\ref{tab:OptimizationBounds}. 

 \begin{table}[t!]
    \centering
    \begin{tabular}{L{0.25\linewidth} L{0.20\linewidth} L{0.17\linewidth} L{0.17\linewidth}}
         \hline
         System \newline Parameter&Classical Inverse \newline Synthesis \cite{Ballester_2022_Application}  &Our \newline Approach (No Geo. Opt.)  &Our \newline Approach (Geo. Opt.) \\
         \hline
         $A$ (eV)& 100 & 90.8 & 90.8\\
         $E_0$ (eV)& 3.73 & 3.71 & 3.71\\
         $E_\text{G}$ (eV)& 1.22 & 1.16 & 1.16\\
         $C$ (eV)& 2.44 & 2.05 & 2.05\\
         $\epsilon_{1 \infty}$ (fixed)& 1 & 1 & 1\\
         $E_\text{C}$ (eV)& 1.75 & 1.70 & 1.70\\
         $d_\text{film}$ ($\mu m$)& 1.122 & 1.118 & 1.118\\
         $\theta_\text{in}$ ($\times10^{-3}$ deg)& 0 & 0 & 0\\
         $c_1$ ($\times10^{-6}$)& 0 & 0 & 2.75\\
         $c_2$ ($\times10^{-3}$ $m^{-1}$)& 0 & 0 & 6.9\\
         RMSE (\%) & \textbf{0.555} & \textbf{0.559} & \textbf{0.559}\\ \hline
    \end{tabular}
    \caption{Comparison of optimization outputs between the state-of-the-art results in \cite{Ballester_2022_Application} and the best results from our proposed approach. Our simulator discovered optical properties that were often close to the solution found with the state-of-the-art method and had RMSE values (used as a figure of merit for optimization) which exhibited a similar fit quality. In addition, our method discovered the same optical properties for two cases where geometry optimization (optimization of surface shapes and beam incidence angle) was disabled and enabled, respectively. This indicates that optimizing for the surface shapes and beam incidence angle may not impact the found material properties, and also implies that our optimizer may have found a set of globally optimum material properties.}
    \label{tab:OptimizationResults}
    \vspace{1em}
\end{table}

 \begin{table}[t!]
    \centering
    \begin{tabular}{L{0.45\linewidth} L{0.20\linewidth} L{0.19\linewidth}}
         \hline
         \centering Simulation setting & ``Rough'' settings for \newline optimization & ``Fine'' \newline settings for refinement\\
         \hline
         Min. Wavelength (nm)& 300 & 300\\
         Max. Wavelength (nm)& 1800 & 1800\\
         Wavelength Spacing (nm)& 2 & 2\\
         Beam Spectral Distr.& Uniform & Uniform\\
         Beam Linewidth (nm)& 2 & 2\\
         \# Points in Spectral Distr. & \textbf{63} & \textbf{255}\\
         Beam Spatial Distribution& Gaussian & Gaussian\\
         Gaussian Beam $\sigma$ (nm)& 40 & 40\\
         Simulation Width (pix) & \textbf{31} & \textbf{255}\\
         Simulation Width (mm)& 0.96 & 0.96\\
         \hline
    \end{tabular}
    \caption{Settings used for validation of our approach. To increase the speed of the optimization process, we first optimized with a set of ``rough'' settings until we found a global optimum. Then we performed a single ``fine'' propagation simulation with the found parameters and different settings to obtain accurate optical property spectra.}
    \label{tab:OptimizationInputs}
\end{table}

\begin{figure}[t!]
\centering\includegraphics[width=\linewidth]{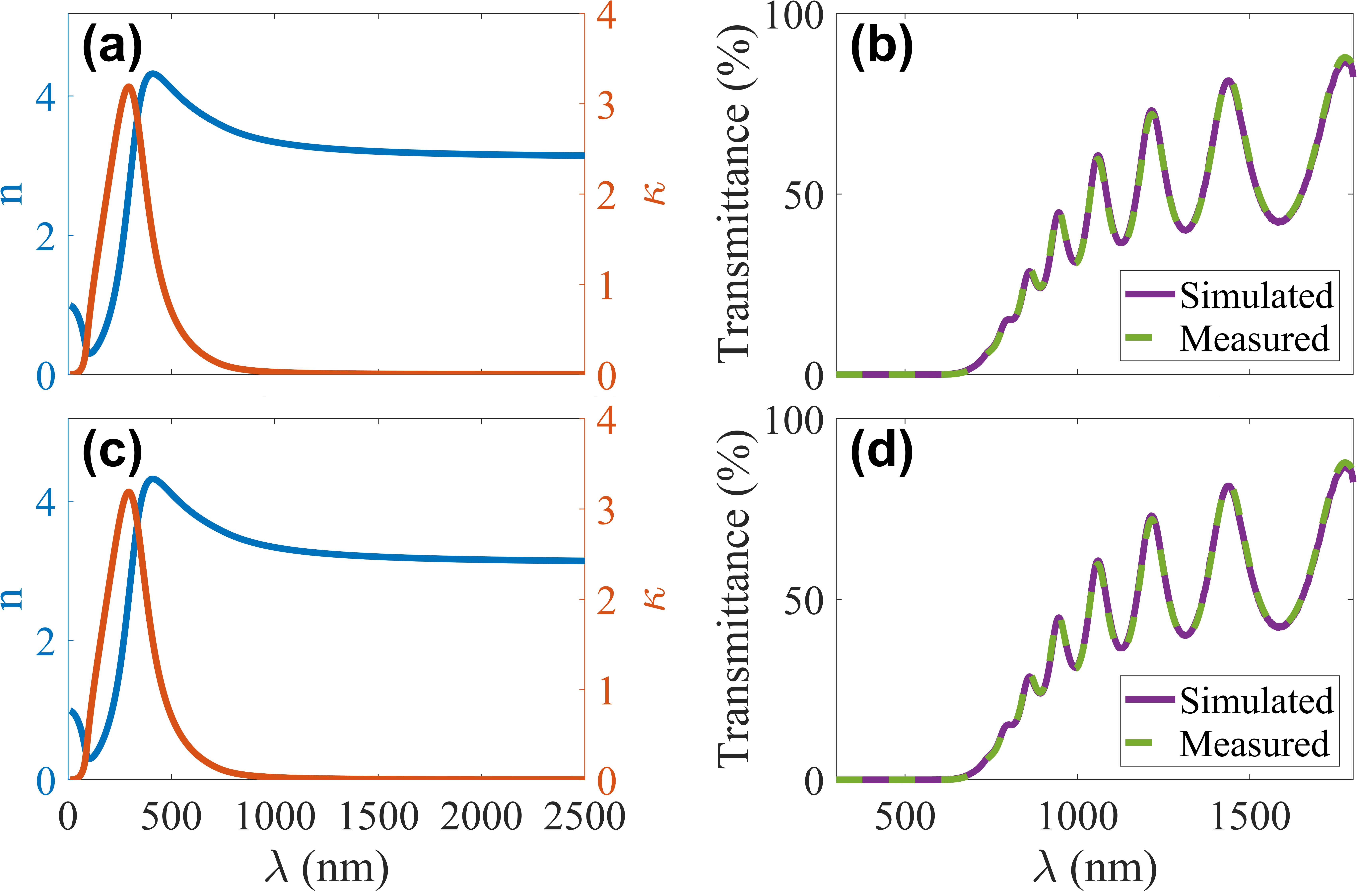}
\caption{(a,c): Computed refractive indices and extinction coefficients using our approach, both without geometry optimization (a), and with geometry optimization~(c).
(b,d): Simulated vs. measured transmittance spectra for our result without geometry optimization (b), and with geometry optimization (d). Both simulated transmittance curves had an RMSE of 0.559\% w.r.t. the measured curves.}
\label{fig:BestFitPlots}
\end{figure}

\begin{table}[t!]
    \centering
    \begin{tabular}{L{0.22\linewidth} L{0.12\linewidth} L{0.115\linewidth} L{0.135\linewidth} L{0.135\linewidth}}
         \hline
         & \multicolumn{2}{c}{No Geometry Opt.} & \multicolumn{2}{c}{Geometry Opt.}\\ \hline
         Compute& \multicolumn{4}{c}{10 nodes with 94 cores for 10 hours each}\\
         Fit Count& \multicolumn{2}{c}{30} & \multicolumn{2}{c}{26}\\ \hline
         & Mean & Std. & Mean & Std. \\ \hline
         $A$ (eV)& 92.7 & 4.0 & 92.5 & 1.4 \\
         $E_0$ (eV)& 3.7032 & 0.008 & 3.7059 & 0.001 \\
         $E_\text{G}$ (eV)& 1.174 & 0.03 & 1.1734 & 0.009\\
         $C$ (eV)& 2.11 & 0.14 & 2.110 & 0.05 \\
         $\epsilon_{1 \infty}$ (fixed)& 1 & 0 & 1 & 0 \\
         $E_\text{C}$ (eV)& 1.713 & 0.03 & 1.7117 & 0.009 \\
         $d_\text{film}$ ($\mu m$)& 1.1186 & 0.0012 & 1.11834 & 0.00011 \\
         $\theta_\text{in}$ ($\times10^{-3}$ deg)& 0 & 0 & 0.001 & 0.005 \\
         $c_1$ ($\times10^{-6}$)& 0 & 0 & 2.49 & 0.2 \\
         $c_2$ ($\times10^{-3}$ $m^{-1}$)& 0 & 0 & -0.3 & 3 \\
         RMSE (\%)& 0.5608 & 0.004 & 0.5597 & 0.004\\ \hline
    \end{tabular}
    \caption{Mean and Standard deviation of evaluated optical properties with and without geometry optimization. When geometry optimization (of incidence angles and thin film surface shape coefficients) is enabled, our proposed approach shows a significant increase in precision by factors between 2.85x and 8.00x.}
    \label{tab:OptimizationStatistics}
    
    \vspace{2em}
    
    \centering
    \begin{tabular}{L{0.25\linewidth} L{0.12\linewidth} L{0.12\linewidth} L{0.12\linewidth} L{0.12\linewidth}}
         \hline
         & \multicolumn{2}{c}{No Geometry Opt.} & \multicolumn{2}{c}{Geometry Opt.} \\ \hline
         System \newline Parameter& Lower bound & Upper bound & Lower bound & Upper bound \\
         \hline
         $A$ (eV)& 70 & 110 & 70 & 110\\
         $E_0$ (eV)& 3.65 & 3.75 & 3.65 & 3.75\\
         $E_\text{G}$ (eV)& 1.1 & 1.35 & 1.1 & 1.35\\
         $C$ (eV)& 1.5 & 3 & 1.5 & 3\\
         $\epsilon_{1 \infty}$ (fixed)& 1 & 1 & 1 & 1\\
         $E_\text{C}$ (eV)& 1.6 & 1.85 & 1.6 & 1.85\\
         $d_\text{film}$ ($\mu m$)& 1.110 & 1.130 & 1.110 & 1.130\\
         $\theta_\text{in}$ ($\times10^{-3}$ deg)& \textbf{0} & \textbf{0} & \textbf{-10} & \textbf{10}\\
         $c_1$ ($\times10^{-6}$)& \textbf{0} & \textbf{0} & \textbf{0} & \textbf{5.2}\\
         $c_2$ ($\times10^{-3}$ $m^{-1}$)& \textbf{0} & \textbf{0} & \textbf{-21.7} & \textbf{21.7}\\
         \hline
    \end{tabular}
    \caption{Global optimization bounds for our approach. We used a set of boundaries that were wide enough to allow for reasonable variations in each material parameter but were tight enough to ensure a fast convergence time and to restrict the amount of local optima that the global optimizer could fall into. For more information on the selection of optimization boundaries, we refer to Appendix \ref{appendix:optimizers}.}
    \label{tab:OptimizationBounds}
\end{table}
From each cycle, we picked the optimization result that produces the smallest RMSE w.r.t. the measured transmittance spectrum, which we respectively consider as the ``best'' result of each cycle. Figure \ref{fig:BestFitPlots} shows the complex refractive index and transmittance spectrum from these two ``best'' results. Table \ref{tab:OptimizationResults} shows a comparison between the final characterization result from \cite{Ballester_2022_Application}, and the ``best'' results of both of our cycles. It can be seen that in both cases, our method produced results with an RMSE that is on par with the state-of-the-art. Moreover, it can be seen that the material parameters evaluated with our method are consistent for both cycles down to multiple significant figures, but are also slightly different from the parameters evaluated in \cite{Ballester_2022_Application}. Since both techniques only use the RMSE of the simulated spectrum w.r.t. the measured spectrum as a figure of merit to evaluate the ``quality'' of the result, it is challenging to say which set of material parameters is closer to the ground truth, and a thorough evaluation would require additional experiments which are outside the scope of this paper.

\begin{figure}[!ht]
\centering\includegraphics[width=\linewidth]{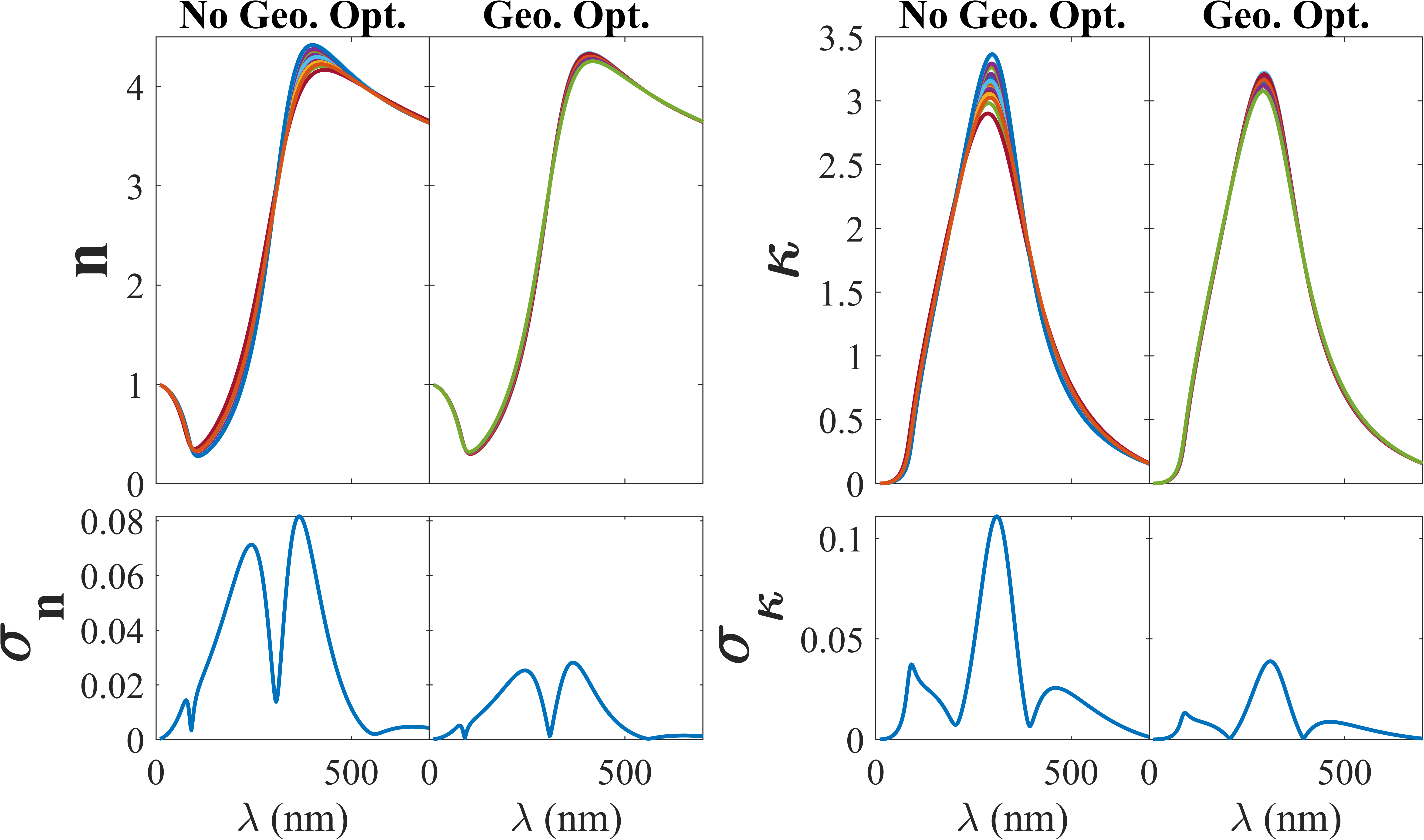}
\caption{(Top) Computed refractive index and extinction coefficient spectra for every single global optimizer fit performed within a fixed time interval (30 without geometry optimization, 26 with geometry optimization). It can be seen that adding geometry optimization significantly increases the precision of fits. (Bottom) Standard deviations of the curves for the top plots.}
\label{fig:ShapeOptImprovement}
\end{figure}

The fact that we perform multiple fits per cycle allows us to additionally evaluate the statistical properties of our method, i.e., calculating the mean value and standard deviation of each parameter. This is shown in Tab. \ref{tab:OptimizationStatistics}. 
It can be seen that the standard deviation values obtained from the optimization cycle with geometry optimization are between 2.8x and 10x smaller than the values from the optimization cycle without geometry optimization. This indicates a clear benefit of the shape optimization method, which leads to significantly more consistent results across different optimizations. We explain the origin of these findings with the nature of global optimizers which can often discover locally optimum solutions, producing similar values of the figure of merit, but have entirely different parameters. Since there is only one \textit{true} set of values for a thin film's optical properties, these locally optimum solutions manifest themselves as statistical variations in the measurement result, arising from the global optimizer itself. By including additional degrees of freedom into the optimization (in our case, the incident angle $\theta_\text{in}$ and surface shape coefficients $c_1, c_2$), the amount of local optimum solutions decreases, reducing this ``optimizer variation'' and thus increasing measurement precision. This improved precision is clearly visible in Fig. \ref{fig:ShapeOptImprovement}. When geometry optimization is disabled, the global optimizer shows large variation for the refractive indices and extinction coefficients at high curvature regions. However, when geometry optimization is enabled, this variability is greatly reduced.

Lastly, we discuss the evaluated geometry optimization results for $c_1$, $c_2$, and $\theta_\text{in}$ which also show evidence that our proposed optimization approach holds merit. As discussed, the characterized sample was previously measured with an atomic force microscope in \cite{Ballester_2022_Application}. These measurements determined that the surface of the measured sample is nearly flat (RMS surface roughness $< 3nm$) in multiple measured regions ranging from  $1 \mu m \times 1 \mu m$ to $5 \mu m \times 5 \mu m$. Although the measured regions do not necessarily coincide with the region in which our spectrophotometer measurement was taken, they are still an indicator that the sample was prepared well without obvious ``irregularities.'' Our optimization-based evaluation results reach the same conclusion: Our evaluated value for $c_1$ (surface tilt) is calculated to $2.75\times10^{-6}$ (equivalent to $0.567 arcsec$ which corresponds to $13.75 pm$ surface variation over $5\upmu$m), with an impressive precision of only $0.2\times10^{-6}$ (equivalent to $0.04 arcsec$). This result implies that our method could be indeed useful for performing surface tilt measurements with absorptive materials.
The beam angle $\theta_\text{in}$ and curvature parameter $c_2$  have both been evaluated to values close to zero with our method. As the sample was nearly flat and the spectrophotometer was calibrated to launch the beam at normal incidence, we believe that these values correctly reflect reality. In the presence of thin films with actual curvature or spectrophotometers with tilted beams, our novel evaluation method might indeed produce a statistically more significant measurement.

\section{Discussion and potential future applications}
Besides the discussed ``classical'' application scenario of inverse synthesis methods, namely the characterization of well-prepared and well-defined thin films, our novel inverse synthesis approach could be applied in a variety of additional situations that are not accounted for in the current state-of-the-art. In particular, this includes characterizing more exotic films and being able to simulate scenarios beyond the surface curvature example that we have demonstrated. In the following, we list a few potential future applications of our novel method:

\noindent \textbf{Multilayer Thin Film Systems:} Our novel method can potentially be used to model and evaluate optical properties of multilayer thin film systems, which are important for the creation of spectral filters, high-reflectance mirrors, antireflection coatings, and other devices \cite{Baumeister_1998_Optical}. As discussed, this would be possible without the need for complicated theoretical analysis with strong approximations and assumptions.

\noindent \textbf{Inhomogenous refractive index distributions:} Our novel method would also be capable of modeling and evaluating changes in the optical properties within a thin film sample. For example, refractive index variations in the longitudinal or transverse direction could be accounted for by modeling the thin film as a combination of multiple smaller thin film layers with different optical properties \cite{vyas_2012}. Another potential application would be to model and evaluate  ``impurities'' of  samples, e.g., if small amounts of a different material are embedded.

\noindent \textbf{3D extension of Inverse Synthesis simulator:}  All of our optimizations thus far have used a 2D model of the system. While this produces precise transmittance spectrum fits for the current sample, the method could be extended to 3D, which potentially even improves the fit accuracy. Moreover, a volumetric 3D representation of the characterized sample would be specifically beneficial for the inhomogeneity characterization discussed above.

\noindent \textbf{Applications beyond thin films:} While we have designed and motivated our method to target applications in optical thin film characterization, the basic idea is universal and not limited to a specific application. This is especially true for the flavor of our method that considers additional parameters, such as surface shape or beam incidence angle. For example, the high precision of the tilt analysis of the sample motivates a potential application of our method in optical surface metrology. Potentially, multi-surface optical systems such as multi-lens systems or optical flats could be characterized. Other possible applications could be found in medical imaging, i.e., for the evaluation and characterization of volumetric multi-layer tissue samples.

We emphasize that virtually all potential future extensions suggested above would require significantly higher computational resources, as the number of variables would need to be increased far beyond what we have demonstrated in this paper. The increased compute comes on top of the fact that, to our knowledge, our current method already takes significantly longer to compute a solution than every existing model. However, as the advancement of modern computers is constantly progressing, we hope that many of the mentioned ideas might become feasible in the near future. \\

While our method in its current state is already capable of producing precise and physically accurate fits for thin film systems, it also still has limitations: The fit quality is currently limited by the material model used for modeling \textit{n} and $\kappa$ (in this paper: TLU model from Foldyna et al. \cite{Foldyna_2004}). For instance, if a thin film sample violates the assumptions of the material model, an accurate fit cannot be achieved. Although this issue is present for classical inverse synthesis methods as well, the integration of a model-less approach like the Swanepoel method \cite{Manifacier_1976, Swanepoel_1983, Swanepoel_1984, jin2017improvement} could yield additional improvements. In addition, the choice of the global optimizer also influences the performance of our method. There exists a large variety of global optimizers still in development
(e.g. bio-inspired heuristic optimizers like the Slime Mold Algorithm \cite{Li_2020}) for different trade-offs between time and accuracy, and being able to avoid local minima. However, global optimizers are generally time-consuming and may not always have a proven convergence to actual global minima for the highly complicated optimization functions seen in thin film transmission spectroscopy. Without fairly tight optimization boundaries (see Tab. \ref{tab:OptimizationBounds}), an inexperienced user would currently still struggle to perform fits in complex situations consisting of multiple unknown surfaces and materials. While geometry optimization reduces the number of local optimum solutions that a global optimizer falls into, our method can still fall into additional local optimum solutions if the optimization bounds are made large while geometry optimization is enabled. Although our experiments have shown that the material model coefficients are not affected over our chosen optimization range, we found during initial attempts at optimization that we could find multiple solutions with strong fits but largely different parabolic curvatures. Again, this can be solved by setting tight limits on the optimization ranges for every considered parameter, as we did in this paper.

\section{Conclusion}
We have demonstrated a novel approach to the inverse-synthesis characterization of semiconductor thin films. Our method is based on an advanced angular spectrum propagation simulator which is paired with a material model and embedded in a global optimizer. We have demonstrated that our approach can precisely model and characterize how light propagates through a medium, with minimal approximations. Our method has been shown to converge to consistent solutions during global optimization, as modeling surface shapes and the beam incidence angle significantly increases the precision of solutions obtained by the optimizer. Finally, our approach has demonstrated effectiveness in modeling the surface tilt and curvature on the top surface of a thin film, and is potentially capable of handling many more scenarios than standard inverse synthesis methods, such as multilayer films and refractive index variation. Despite this increased flexibility of our method, we have demonstrated a thin film characterization with a fit error of only 0.559\% for an amorphous silicon sample, which is on par with current, more rigid, state-of-the-art methods. In the future, we want to expand our method to handle even more complex and ``unusual'' situations and explore novel applications beyond thin film characterization, e.g., in medical imaging.
\\
\\
\\

\appendix
\newpage

\section{Calculation of shape-based reflection and transmission coefficents}
\label{appendix:coefficients}

\noindent
While the Fresnel coefficients of Eq. \ref{eq:FresnelReflectionCoefficient}  are alone sufficient for simulating propagation through a series of parallel planar surfaces, additional adjustments must be made to the reflection simulation at surfaces with varying shapes. If a volume has a surface that is curved outwards, for instance, the portion of the wavefront closer to the center of curvature would travel less far than the portions further away. For small differences in propagation distances less than a wavelength, this effect can be approximated as a phase shift of the respective field and an adjustment of the calculated amount of absorption. Our ``shape-based'' reflection and transmission coefficients apply these phase shifts and intensity changes using a form similar to the Lambert-Beer equation.

First, we define the longitudinal surface shape of a medium as a zero-mean function \textit{S(x)}, which is depicted visually in Fig. \ref{fig:SurfaceFunctionDiagram}(b). For the purposes of calculation, we assume that light is traveling from medium with complex index \(\Tilde{n}_1\) to medium with index \(\Tilde{n}_2\). The phase $\phi{}_{\Delta{}t}(x)$ applied to a transmitted beam, as well as the difference amount of absorption $a_{\Delta{}t}(x)$, is proportional to the optical path difference introduced by the non-planar surface $S(x)$.  Knowing these factors, a ``shape-based'' transmission coefficient $t_{\Delta{}}(x)$ can be derived:

\begin{equation}\label{eq:TransmissivePhaseFromShape}
\phi{}_{\Delta{}t}(x) = k(n_2 - n_1)S(x)
\end{equation}
\begin{equation}\label{eq:TransmissiveAttenuatorFromShape}
a_{\Delta{}t}(x) = k(\kappa_2 - \kappa_1)S(x)
\end{equation}
\begin{equation}\label{eq:ShapeBasedTransmissionCoefficient}
t_{\Delta{}}(x) = \exp{\left(-a_{\Delta{}t}(x) + \iu{}\phi{}_{\Delta{}t}(x))\right)}
\end{equation}

\noindent 
A reflected beam does not travel through the second medium, but it instead travels once through the first medium, and then takes an additional return trip. Thus, the phase and absorption functions applied are not based on index \(n_2\) but instead are entirely based on index \(n_1\). The resulting  ``shape-based'' reflection coefficient is hence derived to:
\begin{equation}\label{eq:ReflectivePhaseFromShape}
\phi{}_{\Delta{}r}(x) = -2kn_1S(x)
\end{equation}
\begin{equation}\label{eq:ReflectiveAttenuatorFromShape}
a_{\Delta{}r}(x) = -2k\kappa_1S(x)
\end{equation}
\begin{equation}\label{eq:ShapeBasedReflectionCoefficient}
r_{\Delta{}}(x) = \exp{\left(-a_{\Delta{}r}(x) + \iu{}\phi{}_{\Delta{}r}(x))\right)}
\end{equation}

\section{Thin film material model}
\label{appendix:tluModel}

\noindent
The thin film material model used in this work consists of a Single-Oscillator Tauch-Lorentz-Urbach (TLU) model developed by Foldyna et. al \cite{Foldyna_2004}, which generates spectra for the real and imaginary parts of the refractive index from a list of six energies. In particular, we can consider the complex dielectric function (relative permittivity) $\Tilde{\epsilon} = \epsilon_1 + \iu{} \epsilon_2$. The TLU model defines the Urbach energy and amplitude as follows:

\begin{subequations}\label{eq:main}
\begin{align}
E_\text{u} &= \frac{E_\text{C}-E_\text{G}}{ 2-2E_\text{C}(E_\text{C}-E_\text{G})  \frac{ C^2+2(E_\text{C}^2-E_0^2) }{C^2 E_\text{C}^2+(E_\text{C}^2-E_0^2)^2}} \tag{\ref*{eq:main}.a},
\\ 
A_\text{u} &= \exp{\left(-\frac{E_\text{C}}{E_\text{u}}\right)} \frac{A E_0 C (E_\text{C}-E_\text{G})^2}{(E_\text{C}^2-E_0^2)^2+C^2 E_\text{C}^2} \tag{\ref*{eq:main}.b}.
\end{align}
\end{subequations}

Then, the imaginary part of the dielectric function is

\begin{equation} \label{eq:TLU}
\epsilon_2(E) = 
\begin{cases} 
\frac{A_{\text{u}}}{E} \exp{\left( \frac{E}{E_{\text{u}}}\right)}, & \text{for } 0 \leq E < E_{\text{C}} \\
\frac{1}{E}\frac{A E_0 C (E-E_\text{G})^2}{ (E^2-E_0^2)^2 +C^2 E^2}, & \text{for } E \geq E_\text{C}
\end{cases}
\end{equation}

For lower energies, we see the empirical Urbach exponential model \cite{urbach1953long, guerra2016urbach}, related to the level of disorder of the material, while for higher energies, we have the theoretical Tauc-Lorentz oscillator model \cite{tauc1966optical, jellison1996parameterization}. The real part $\epsilon_1$ can then be calculated analytically from $\epsilon_2$ using the Kramers-Kronig relation \cite{larruquert2020optical, Marquez_2021}. Considering the complex refractive index $\Tilde{n} = n + \iu{}\kappa$, we have the relation $\Tilde{n}^2 = \Tilde{\epsilon} = \epsilon_1 + i \epsilon_2$ for non-magnetic materials. Therefore, we can now express both properties $n(\lambda)$ and $\kappa(\lambda)$ with a set of a few real coefficients using this advanced dispersion model. Finding such parameters is enough to optically characterize the thin film.

\section{Local and global optimizers}
\label{appendix:optimizers}

We note that local optimizers, and some global optimizers, struggle to determine surface shapes when initialized close to the globally optimum solution. We observed this when testing the commonly-used Simulated Annealing \cite{Henderson_2003_Theory} and Nelder Mead \cite{Lagarias_1998_Convergence} optimizers to see if they could recover a geometry-optimized solution using a non-geometry-optimized fit as an initial condition. We found that both optimizers struggled to leave the local minimum of the non-geometry-optimized starting point, and would always report completely flat surface shapes, and normal-incidence beam angles, as a result. Thus, we suggest that to perform our method, a user should first perform a genetic algorithm fit \cite{Alhijawi_2024_Genetic} without geometry optimization enabled to get close to true material parameters. The genetic algorithm tended to consistently find strong fits even with wide bounds for the material properties, making it a good choice for an initial fit. Then, the original fit results should be used to define a set of boundaries for a second genetic algorithm fit, with geometry optimization enabled. This second fit has reduced measurement variance due to the geometry optimization and thus provides precise material parameters.

\section*{Bibliography}
\bibliographystyle{ieeetr}
\bibliography{references}

\begin{thebibliography}{10}

\bibitem{Chayma_2020}
C.~Abed, S.~Fernández, and H.~Elhouichet, ``Studies of optical properties of zno:mgo thin films fabricated by sputtering from home-made stable oversize targets,'' {\em Optik}, vol.~216, p.~164934, 05 2020.

\bibitem{Rodriguez-Tapiador_2023}
M.~I. Rodríguez-Tapiador, J.~M. Asensi, M.~Roldán, J.~Merino, J.~Bertomeu, and S.~Fernández, ``Copper nitride: A versatile semiconductor with great potential for next-generation photovoltaics,'' {\em Coatings}, vol.~13, no.~6, 2023.

\bibitem{lopezenhanced}
A.~J. Lopez-Garcia, G.~Alvarez-Suarez, E.~Ros, P.~Ortega, C.~Voz, J.~Puigdollers, and A.~P. Rodr{\'\i}guez, ``Enhanced selective contact behavior in a-si: H/oxide transparent pv devices via dipole layer integration,'' {\em Solar RRL}, 2024.

\bibitem{marquez2023complex}
E.~Marquez, M.~Ballester, M.~Garcia, M.~Cintado, A.~Marquez, J.~Ruiz, S.~Fern{\'a}ndez, E.~Blanco, F.~Willomitzer, and A.~Katsaggelos, ``Complex dielectric function of h-free a-si films: Photovoltaic light absorber,'' {\em Materials Letters}, vol.~345, p.~134485, 2023.

\bibitem{petrik2012optical}
P.~Petrik, ``Optical thin film metrology for optoelectronics,'' in {\em Journal of Physics: Conference Series}, vol.~398, p.~012002, IOP Publishing, 2012.

\bibitem{islam2020band}
M.~N. Islam, J.~Podder, K.~S. Hossain, and S.~Sagadevan, ``Band gap tuning of p-type al-doped tio2 thin films for gas sensing applications,'' {\em Thin Solid Films}, vol.~714, p.~138382, 2020.

\bibitem{panda2012preparation}
S.~Panda and C.~Jacob, ``Preparation of transparent zno thin films and their application in uv sensor devices,'' {\em Solid-State Electronics}, vol.~73, pp.~44--50, 2012.

\bibitem{nesheva2019changes}
D.~Nesheva, P.~Petrik, T.~Hristova-Vasileva, Z.~Fogarassy, B.~Kalas, M.~{\v{S}}{\'c}epanovi{\'c}, S.~Kaschieva, S.~N. Dmitriev, and K.~Antonova, ``Changes in composite nc-si-sio2 thin films caused by 20 mev electron irradiation,'' {\em Nuclear Instruments and Methods in Physics Research Section B: Beam Interactions with Materials and Atoms}, vol.~458, pp.~159--163, 2019.

\bibitem{marquez2023optical}
E.~M{\'a}rquez, E.~Blanco, M.~Garc{\'\i}a-Gurrea, M.~Cintado~Puerta, M.~Dom{\'\i}nguez de~la Vega, M.~Ballester, J.~M{\'a}nuel, M.~Rodr{\'\i}guez-Tapiador, and S.~Fern{\'a}ndez, ``Optical properties of reactive rf magnetron sputtered polycrystalline cu3n thin films determined by uv/visible/nir spectroscopic ellipsometry: An eco-friendly solar light absorber,'' {\em Coatings}, vol.~13, no.~7, p.~1148, 2023.

\bibitem{kumar2014evolution}
R.~A. Kumar, P.~C. Lekha, C.~Sanjeeviraja, and D.~P. Padiyan, ``Evolution of structural disorder using raman spectra and urbach energy in gese0. 5s1. 5 thin films,'' {\em Journal of non-crystalline solids}, vol.~405, pp.~21--26, 2014.

\bibitem{ballester2022energy}
M.~Ballester, A.~M{\'a}rquez, C.~Garc{\'\i}a-V{\'a}zquez, J.~D{\'\i}az, E.~Blanco, D.~Minkov, S.~Fern{\'a}ndez-Ruano, F.~Willomitzer, O.~Cossairt, and E.~M{\'a}rquez, ``Energy-band-structure calculation by below-band-gap spectrophotometry in thin layers of non-crystalline semiconductors: A case study of unhydrogenated a-si,'' {\em Journal of Non-Crystalline Solids}, vol.~594, p.~121803, 2022.

\bibitem{amato2020observation}
A.~Amato, S.~Terreni, M.~Granata, C.~Michel, B.~Sassolas, L.~Pinard, M.~Canepa, and G.~Cagnoli, ``Observation of a correlation between internal friction and urbach energy in amorphous oxides thin films,'' {\em Scientific Reports}, vol.~10, no.~1, p.~1670, 2020.

\bibitem{Ballester_2022_Application}
M.~Ballester, M.~García, A.~P. Márquez, E.~Blanco, S.~M. Fernández, D.~Minkov, A.~K. Katsaggelos, O.~Cossairt, F.~Willomitzer, and E.~Márquez, ``Application of the holomorphic tauc-lorentz-urbach function to extract the optical constants of amorphous semiconductor thin films,'' {\em Coatings}, vol.~12, no.~10, 2022.

\bibitem{Kreis_2022_Foundations}
T.~Kreis, ``Foundations of optical interferometry,'' in {\em Full Field Optical Metrology and Applications}, 2053-2563, pp.~1--1 to 1--19, IOP Publishing, 2022.

\bibitem{Thompkins_2005}
H.~G. Tompkins and E.~A. Irene, eds., {\em Handbook of Ellipsometry}.
\newblock Springer Berlin and William Andrew Publishing, 2005.

\bibitem{fujiwara2007spectroscopic}
H.~Fujiwara, {\em Spectroscopic ellipsometry: principles and applications}.
\newblock John Wiley \& Sons, 2007.

\bibitem{kalas2017ellipsometric}
B.~Kalas, B.~Pollakowski, A.~Nutsch, C.~Streeck, J.~Nador, M.~Fried, B.~Beckhoff, and P.~Petrik, ``Ellipsometric and x-ray spectrometric investigation of fibrinogen protein layers,'' {\em physica status solidi c}, vol.~14, no.~12, p.~1700210, 2017.

\bibitem{romanenko2020membrane}
A.~Romanenko, B.~Kalas, P.~Hermann, O.~Hakkel, L.~Ill{\'e}s, M.~Fried, P.~F{\"u}rjes, G.~Gyulai, and P.~Petrik, ``Membrane-based in situ mid-infrared spectroscopic ellipsometry: A study on the membrane affinity of polylactide-co-glycolide nanoparticulate systems,'' {\em Analytical Chemistry}, vol.~93, no.~2, pp.~981--991, 2020.

\bibitem{lohner2020determination}
T.~Lohner, E.~Szil{\'a}gyi, Z.~Zolnai, A.~N{\'e}meth, Z.~Fogarassy, L.~Ill{\'e}s, E.~K{\'o}tai, P.~Petrik, and M.~Fried, ``Determination of the complex dielectric function of ion-implanted amorphous germanium by spectroscopic ellipsometry,'' {\em Coatings}, vol.~10, no.~5, p.~480, 2020.

\bibitem{marquez2023mid}
E.~M{\'a}rquez, E.~Blanco, J.~M. M{\'a}nuel, M.~Ballester, M.~Garc{\'\i}a-Gurrea, M.~I. Rodr{\'\i}guez-Tapiador, S.~M. Fern{\'a}ndez, F.~Willomitzer, and A.~K. Katsaggelos, ``Mid-infrared (mir) complex refractive index spectra of polycrystalline copper-nitride films by ir-vase ellipsometry and their fib-sem porosity,'' {\em Coatings}, vol.~14, no.~1, p.~5, 2023.

\bibitem{Heavens_1991}
O.~S. Heavens, {\em Optical properties of thin solid films}.
\newblock Courier Corporation, 1991.

\bibitem{Bosch_1998}
S.~Bosch, ``{Spectrophotometry, ellipsometry, and computer simulation in thin film developments},'' in {\em ROMOPTO '97: Fifth Conference on Optics} (V.~I. Vlad and D.~C. Dumitras, eds.), vol.~3405, pp.~1120 -- 1131, International Society for Optics and Photonics, SPIE, 1998.

\bibitem{van2005determination}
R.~Van~Veen, H.~J. Sterenborg, A.~Pifferi, A.~Torricelli, E.~Chikoidze, and R.~Cubeddu, ``Determination of visible near-ir absorption coefficients of mammalian fat using time-and spatially resolved diffuse reflectance and transmission spectroscopy,'' {\em Journal of biomedical optics}, vol.~10, no.~5, pp.~054004--054004, 2005.

\bibitem{saleh2017evaluation}
M.~H. Saleh, N.~M. Ershaidat, M.~J.~A. Ahmad, B.~N. Bulos, and M.~M. A.-G. Jafar, ``Evaluation of spectral dispersion of optical constants of a-se films from their normal-incidence transmittance spectra using swanepoel algebraic envelope approach,'' {\em Optical Review}, vol.~24, pp.~260--277, 2017.

\bibitem{minkov2020perfecting}
D.~Minkov, G.~Angelov, R.~Nestorov, and E.~Marquez, ``Perfecting the dispersion model free characterization of a thin film on a substrate specimen from its normal incidence interference transmittance spectrum,'' {\em Thin Solid Films}, vol.~706, p.~137984, 2020.

\bibitem{Hedderich_2023}
H.~G. Hedderich, {\em Cary 6000i UV-Vis-NIR --- User’s Guide}, October 2023.

\bibitem{PerkinElmer_2016}
PerkinElmer, {\em HIGH-PERFORMANCE LAMBDA SPECTROMETERS}, 2016.

\bibitem{Manifacier_1976}
J.~C. Manifacier, J.~Gasiot, and J.~P. Fillard, ``A simple method for the determination of the optical constants n, k and the thickness of a weakly absorbing thin film,'' {\em Journal of Physics E: Scientific Instruments}, vol.~9, p.~1002, nov 1976.

\bibitem{Swanepoel_1983}
R.~Swanepoel, ``Determination of the thickness and optical constants of amorphous silicon,'' {\em Journal of Physics E: Scientific Instruments}, vol.~16, p.~1214, dec 1983.

\bibitem{Swanepoel_1984}
R.~Swanepoel, ``Determination of surface roughness and optical constants of inhomogeneous amorphous silicon films,'' {\em Journal of Physics E: Scientific Instruments}, vol.~17, p.~896, oct 1984.

\bibitem{jin2017improvement}
Y.~Jin, B.~Song, Z.~Jia, Y.~Zhang, C.~Lin, X.~Wang, and S.~Dai, ``Improvement of swanepoel method for deriving the thickness and the optical properties of chalcogenide thin films,'' {\em Optics express}, vol.~25, no.~1, pp.~440--451, 2017.

\bibitem{marquez2019influence}
E.~M{\'a}rquez, E.~Saugar, J.~D{\'\i}az, C.~Garc{\'\i}a-V{\'a}zquez, S.~Fern{\'a}ndez-Ruano, E.~Blanco, J.~Ruiz-P{\'e}rez, and D.~Minkov, ``The influence of ar pressure on the structure and optical properties of non-hydrogenated a-si thin films grown by rf magnetron sputtering onto room-temperature glass substrates,'' {\em Journal of Non-Crystalline Solids}, vol.~517, pp.~32--43, 2019.

\bibitem{tejada2019determination}
A.~Tejada, L.~Monta{\~n}ez, C.~Torres, P.~Llontop, L.~Flores, F.~De~Zela, A.~Winnacker, and J.~Guerra, ``Determination of the fundamental absorption and optical bandgap of dielectric thin films from single optical transmittance measurements,'' {\em Applied Optics}, vol.~58, no.~35, pp.~9585--9594, 2019.

\bibitem{Ballester_2022_Comparison}
M.~Ballester, A.~P. M\'{a}rquez, S.~Banerjee, J.~J. Ru\'{i}z-P\'{e}rez, O.~Cossairt, A.~K. Katsaggelos, F.~Willomitzer, and E.~M\'{a}rquez, ``Comparison of optical characterization methods for transmission spectroscopy,'' in {\em Imaging and Applied Optics Congress 2022 (3D, AOA, COSI, ISA, pcAOP)}, p.~JW5D.4, Optica Publishing Group, 2022.

\bibitem{Dobrowolski_1983}
J.~A. Dobrowolski, F.~C. Ho, and A.~Waldorf, ``Determination of optical constants of thin film coating materials based on inverse synthesis,'' {\em Appl. Opt.}, vol.~22, pp.~3191--3200, Oct 1983.

\bibitem{Poelman_2003}
D.~Poelman and P.~F. Smet, ``Methods for the determination of the optical constants of thin films from single transmission measurements: a critical review,'' {\em Journal of Physics D: Applied Physics}, vol.~36, p.~1850, jul 2003.

\bibitem{ventura2005optimization}
S.~Ventura, E.~Birgin, J.~M. Martinez, and I.~Chambouleyron, ``Optimization techniques for the estimation of the thickness and the optical parameters of thin films using reflectance data,'' {\em Journal of Applied Physics}, vol.~97, no.~4, 2005.

\bibitem{Knittl_1976}
Z.~Knittl and T.~Z. Knittl., {\em Optics of thin films; an optical multilayer theory}.
\newblock John Wiley \& Sons, Inc., 1976.

\bibitem{yeh2006optical}
P.~Yeh, {\em Optical Waves in Layered Media}.
\newblock Wiley Online Library, 2006.

\bibitem{prentice2000coherent}
J.~Prentice, ``Coherent, partially coherent and incoherent light absorption in thin-film multilayer structures,'' {\em Journal of Physics D: Applied Physics}, vol.~33, no.~24, p.~3139, 2000.

\bibitem{Ruiz_2020}
J.~J. Ruiz-Pérez and E.~M. Navarro, ``Optical transmittance for strongly-wedge-shaped semiconductor films: Appearance of envelope-crossover points in amorphous as-based chalcogenide materials,'' {\em Coatings}, vol.~10, no.~11, 2020.

\bibitem{Alhijawi_2024_Genetic}
B.~Alhijawi and A.~Awajan, ``Genetic algorithms: theory, genetic operators, solutions, and applications,'' {\em Evolutionary Intelligence}, vol.~17, pp.~1245--1256, Feb 2023.

\bibitem{Goodman_1969}
J.~W. Goodman, {\em Introduction to Fourier optics}.
\newblock McGraw-Hill, 1969.

\bibitem{Bass_2024_Angular}
J.~M. Bass, M.~Ballester, S.~M. Fernández, A.~K. Katsaggelos, E.~Márquez, and F.~Willomitzer, ``Angular spectrum approach to optical characterization of thin film materials using transmission spectroscopy,'' in {\em Proceedings of the 14th International Conference on Optics-photonics Design and Fabrication (ODF)}, 2024.

\bibitem{Bass_2024_Increasing}
J.~M. Bass, M.~Ballester, S.~M. Fernández, A.~K. Katsaggelos, E.~Márquez, and F.~Willomitzer, ``Increasing the precision of transmission spectroscopy by optimization of thin film surface shapes,'' in {\em Optica Open}, 2024.

\bibitem{Sherman_1967_Application}
G.~C. Sherman, ``Application of the convolution theorem to rayleigh's integral formulas,'' {\em J. Opt. Soc. Am.}, vol.~57, pp.~546--547, Apr 1967.

\bibitem{vyas_2012}
U.~Vyas and D.~Christensen, ``Ultrasound beam simulations in inhomogeneous tissue geometries using the hybrid angular spectrum method,'' {\em IEEE Transactions on Ultrasonics, Ferroelectrics, and Frequency Control}, vol.~59, pp.~1093--1100, June 2012.
\newblock Conference Name: IEEE Transactions on Ultrasonics, Ferroelectrics, and Frequency Control.

\bibitem{Xu_2022}
R.~Xu, M.~Feng, Z.~Chen, J.~Yang, D.~Han, J.~Xie, and F.~Song, ``Non-uniform angular spectrum method in a complex medium based on iteration,'' {\em Opt. Lett.}, vol.~47, pp.~1972--1975, Apr 2022.

\bibitem{Mellin_2001}
S.~D. Mellin and G.~P. Nordin, ``Limits of scalar diffraction theory and an iterative angular spectrum algorithm for finite aperture diffractive optical element design,'' {\em Opt. Express}, vol.~8, pp.~705--722, Jun 2001.

\bibitem{king1999analysis}
R.~L. King, C.~Ruffin, F.~LaMastus, and D.~Shaw, ``The analysis of hyperspectral data using savitzky-golay filtering-practical issues. 2,'' in {\em IEEE 1999 International Geoscience and Remote Sensing Symposium. IGARSS'99 (Cat. No. 99CH36293)}, vol.~1, pp.~398--400, IEEE, 1999.

\bibitem{zimmermann2013optimizing}
B.~Zimmermann and A.~Kohler, ``Optimizing savitzky--golay parameters for improving spectral resolution and quantification in infrared spectroscopy,'' {\em Applied spectroscopy}, vol.~67, no.~8, pp.~892--902, 2013.

\bibitem{zhao2014parameters}
A.-X. Zhao, X.-J. Tang, Z.-H. Zhang, and J.-H. Liu, ``The parameters optimization selection of savitzky-golay filter and its application in smoothing pretreatment for ftir spectra,'' in {\em 2014 9th IEEE Conference on industrial electronics and applications}, pp.~516--521, IEEE, 2014.

\bibitem{Foldyna_2004}
M.~Foldyna, K.~Postava, J.~Bouchala, J.~Pistora, and T.~Yamaguchi, ``{Model dielectric functional of amorphous materials including Urbach tail},'' in {\em Microwave and Optical Technology 2003} (J.~Pistora, K.~Postava, M.~Hrabovsky, and B.~S. Rawat, eds.), vol.~5445, pp.~301 -- 305, International Society for Optics and Photonics, SPIE, 2004.

\bibitem{Anders_2002_Interior}
A.~Forsgren, P.~E. Gill, and M.~H. Wright, ``Interior methods for nonlinear optimization,'' {\em SIAM Review}, vol.~44, no.~4, pp.~525--597, 2002.

\bibitem{Gondzio_2012_Interior}
J.~Gondzio, ``Interior point methods 25 years later,'' {\em European Journal of Operational Research}, vol.~218, no.~3, pp.~587--601, 2012.

\bibitem{Fernandez_2021_Sputtered}
S.~Fernández, J.~J. Gandía, E.~Saugar, M.~B. Gómez-Mancebo, D.~Canteli, and C.~Molpeceres, ``Sputtered non-hydrogenated amorphous silicon as alternative absorber for silicon photovoltaic technology,'' {\em Materials}, vol.~14, no.~21, 2021.

\bibitem{Baumeister_1998_Optical}
P.~W. Baumeister, {\em Optical coating technology}.
\newblock SPIE Publications, 1998.

\bibitem{Li_2020}
S.~Li, H.~Chen, M.~Wang, and A.~A. Heidari, ``Slime mould algorithm: A new method for stochastic optimization,'' {\em Future Generation Computer Systems}, pp.~300--323, 04 2020.

\bibitem{urbach1953long}
F.~Urbach, ``The long-wavelength edge of photographic sensitivity and of the electronic absorption of solids,'' {\em Physical review}, vol.~92, no.~5, p.~1324, 1953.

\bibitem{guerra2016urbach}
J.~Guerra, J.~Angulo, S.~Gomez, J.~Llamoza, L.~Monta{\~n}ez, A.~Tejada, J.~T{\"o}fflinger, A.~Winnacker, and R.~Weing{\"a}rtner, ``The urbach focus and optical properties of amorphous hydrogenated sic thin films,'' {\em Journal of Physics D: Applied Physics}, vol.~49, no.~19, p.~195102, 2016.

\bibitem{tauc1966optical}
J.~Tauc, R.~Grigorovici, and A.~Vancu, ``Optical properties and electronic structure of amorphous germanium,'' {\em physica status solidi (b)}, vol.~15, no.~2, pp.~627--637, 1966.

\bibitem{jellison1996parameterization}
G.~Jellison~Jr and F.~Modine, ``Parameterization of the optical functions of amorphous materials in the interband region,'' {\em Applied Physics Letters}, vol.~69, no.~3, pp.~371--373, 1996.

\bibitem{larruquert2020optical}
J.~I. Larruquert and L.~V.~R. de~Marcos, ``Optical constants at complex energies: local deconvolution,'' {\em Optics Express}, vol.~28, no.~8, pp.~11767--11779, 2020.

\bibitem{Marquez_2021}
E.~Márquez, J.~J. Ruíz-Pérez, M.~Ballester, A.~P. Márquez, E.~Blanco, D.~Minkov, S.~M.~F. Ruano, and E.~Saugar, ``Optical characterization of h-free a-si layers grown by rf-magnetron sputtering by inverse synthesis using matlab: Tauc–lorentz–urbach parameterization,'' {\em Coatings}, vol.~11, no.~11, 2021.

\bibitem{Henderson_2003_Theory}
D.~Henderson, S.~H. Jacobson, and A.~W. Johnson, ``The theory and practice of simulated annealing,'' in {\em Handbook of Metaheuristics} (F.~Glover and G.~A. Kochenberger, eds.), pp.~287--319, Boston, MA: Springer US, 2003.

\bibitem{Lagarias_1998_Convergence}
J.~C. Lagarias, J.~A. Reeds, M.~H. Wright, and P.~E. Wright, ``Convergence properties of the nelder--mead simplex method in low dimensions,'' {\em SIAM Journal on Optimization}, vol.~9, no.~1, pp.~112--147, 1998.

\end{thebibliography}

\end{document}